\documentclass[12pt]{article}

\usepackage{bbm}
\usepackage{epsfig}
\usepackage{array}
\usepackage{float}
\usepackage{dsfont}
\usepackage{amstext}
\usepackage{amsmath}
\usepackage{amssymb}
\usepackage{amsfonts}   
\usepackage{graphicx}
\usepackage{color}
\usepackage{cancel}
\usepackage{fancybox}
\usepackage{amsmath, amssymb, graphics}
\usepackage{rotating}
\usepackage{a4}
\usepackage{a4wide}

\newcommand{\bav}{\begin{array}{cccc}}
\newcommand{\bas}{\begin{array}{cccccc}}
\newcommand{\mathsym}[1]{{}}

\newcommand{\baz}{\begin{array}{cc}}
\newcommand{\bad}{\begin{array}{ccc}}
\newcommand{\bi}{\begin{itemize}}
\newcommand{\ei}{\end{itemize}}
\newcommand{\ba}{\begin{array}{c}}
\newcommand{\ea}{\end{array}}

\newcommand{\dms}{\mbox{$\Delta m^2_{\odot}$}}
\newcommand{\dma}{\mbox{$\Delta m^2_{\rm A}$}}

\def\be{\begin{equation}}
\def\ee{\end{equation}}
\newcommand{\bea}{\begin{equation} \begin{array}{c}}
\newcommand{\eea}{ \end{array} \end{equation}}

\def\gs{\mathrel{
   \rlap{\raise 0.511ex \hbox{$>$}}{\lower 0.511ex \hbox{$\sim$}}}}
\def\ls{\mathrel{
   \rlap{\raise 0.511ex \hbox{$<$}}{\lower 0.511ex \hbox{$\sim$}}}}

\numberwithin{equation}{section}

\begin{document}

\title{
\vskip 0.4cm
\Large \bf
Phenomenological Consequences of sub-leading Terms in See-Saw Formulas
}\vspace{2.5cm}
\author{
Hans Hettmansperger\thanks{email: \tt
hhettman@googlemail.com}~,~~
Manfred Lindner\thanks{email: \tt 
manfred.lindner@mpi-hd.mpg.de}~,~~
Werner Rodejohann\thanks{email: \tt
werner.rodejohann@mpi-hd.mpg.de}  \\ \\
{\normalsize \it Max--Planck--Institut f\"ur Kernphysik,}\\
{\normalsize \it  Postfach 103980, D--69029 Heidelberg, Germany} }
\date{}
\maketitle
\thispagestyle{empty}
\vspace{-0.8cm}
\begin{abstract}
\noindent
Several aspects of next-to-leading (NLO) order corrections 
to see-saw formulas are discussed and phenomenologically relevant
situations are identified. 
We generalize the formalism to calculate the NLO terms developed for the 
type I see-saw to variants like the inverse, double or  
linear see-saw, i.e., to cases in which more than two mass scales are
present. In the standard type I
case with very heavy fermion singlets the sub-leading terms are
negligible. However, effects in the percent regime are possible 
when sub-matrices of the complete neutral fermion
mass matrix obey a moderate hierarchy, e.g.~weak scale and TeV
scale. Examples are 
cancellations of large terms leading to small neutrino masses, 
or inverse see-saw scenarios. 
We furthermore identify
situations in which no NLO corrections to certain observables arise, 
namely for $\mu$--$\tau$ symmetry and cases with a vanishing neutrino mass. 
Finally, we emphasize that the unavoidable unitarity violation in 
see-saw scenarios with extra fermions can be calculated with the formalism in a
straightforward manner. 
 
\end{abstract}

\newpage

\section{\label{sec:intro}Introduction}
Neutrino masses are small. This simple fact is usually attributed to
the presence of a see-saw mechanism. In its simplest and most often
studied manifestation, the type I see-saw \cite{seesaw1}, 
the low energy Majorana neutrino mass matrix is 
\[ 
m_\nu = - m_D^T \, M_R^{-1} \, m_D \, , 
\]
where $m_D$ is a Dirac and $M_R$ a Majorana mass matrix. The above
relation is formally obtained in the limit of ``$M_R \gg m_D$'', which
means that the eigenvalues of $M_R$ are much larger than the entries
of $m_D$. While one usually gives the expression for $m_\nu$ with a
equality sign ``$=$'', it should strictly speaking be an approximative equal
sign ``$\simeq$'', as there are higher order, or next-to-leading order (NLO), terms
which correct it. In the standard type I see-saw with heavy fermion
singlets the corrections are usually negligible. 
In the present paper we will study the NLO corrections in detail and
identify phenomenologically relevant applications. 
A formalism to give those terms at arbitrary order 
has been developed by Grimus and Lavoura in Ref.~\cite{GL}. 
We generalize that formalism\footnote{A similar Ansatz for the unitary matrix 
diagonalizing the full neutral fermion mass matrix has been 
proposed in Ref.~\cite{Jose}. The same results could be obtained
with this approach.} to several of the see-saw variants 
which are discussed in the 
literature, such as the double \cite{double}, inverse \cite{inverse}, 
linear \cite{linear} or singular \cite{sing0} see-saw. 
We note that in scenarios in which the see-saw scale is lowered to TeV
scale, 
NLO corrections in the percent regime are 
possible, which in the light of up-coming neutrino precision
experiments is surely not negligible\footnote{We will
discuss here only the corrections to the low energy mass matrix $m_\nu$ and
leave consequences for the heavy singlets for further study.}. 
This occurs for instance in scenarios in which cancellations of large
terms lead to small neutrino mass, or in inverse see-saw frameworks. 
These cases have in common that sub-matrices of the complete neutral fermion
mass matrix obey a moderate hierarchy. 

One common aspect of all see-saw mechanisms with additional fermions
is the violation of unitarity of the $3\times3$ mixing matrix which 
describes the mixing of the three active neutrinos.  
The formalism to calculate NLO see-saw
corrections is applicable to calculate the magnitude and structure of
those. This allows to obtain formulae for unitarity violation in a 
simple and straightforward manner for the see-saw variants. 

We furthermore identify situations in which no corrections to 
certain parameters arise. We will
show that in case of $\mu$--$\tau$ symmetry there are no NLO corrections to
$U_{e3} = 0$ and $\theta_{23} = \pi/4$. Another finding is that if 
one neutrino is massless, then the mixing matrix elements 
associated with this massless state receive no corrections. 
The massless neutrino does not mix with the heavy ones.\\

The paper is build up as follows: in Section \ref{sec:GL} 
the formalism to calculate higher order corrections to the type I
see-saw term is reviewed. The connection to the inherent unitarity
violation is noted. In Section \ref{sec:exa} we discuss examples 
on the application of the NLO terms, and identify cases which are
stable. In Section \ref{sec:var} we show how the formalism can be
applied to see-saw variants, before we conclude in Section
\ref{sec:concl}.

\section{NLO Terms to the Type I See-Saw Mechanism}
\label{sec:GL}
In what follows we will review the derivation of the NLO terms to the
type I see-saw formula. The reader familiar with it can continue in
Section \ref{sec:app}, where some of its applications are studied. 

\subsection{\label{sec:}Derivation of NLO Terms}

The conventional type I see-saw mechanism \cite{seesaw1} contains
after electroweak symmetry breaking two mass terms in the Lagrangian: 
\be \label{eq:I}
{\cal L} = \overline{N_R} \, m_D \, \nu_L + \frac 12 \, 
\overline{N_R} \, M_R \, N_R^c + h.c.
\ee
Here $\nu_L$ are left-handed neutrinos, $N_R$ are right-handed
singlets, and $m_D$ ($M_R$) is the Dirac (Majorana) mass matrix. 
The effective mass matrix at low energy is conventionally obtained by
integrating out the heavy states $N_R$, and the result is 
\be
m_\nu \simeq - m_D^T \, M_R^{-1} \, m_D \, .
\ee 
The approximative nature of this expression is noteworthy. Formally,
the effective mass matrix is obtained by diagonalizing the total 
mass matrix in the basis $(\nu_L^c, N_R)$:
\be\label{eqn:typeIMass}
{\cal M} = 
\left( 
\baz 
0 & m_D^T \\
m_D & M_R 
\ea 
\right) .
\ee
We note here for later use that the inverse of a matrix of this
texture is 
\be \label{eq:Mi}
{\cal M}^{-1} = 
\left( 
\baz
- m_D^{-1} \, M_R \, (m_D^T)^{-1} & m_D^{-1} \\
(m_D^T)^{-1} & 0 
\ea 
\right). 
\ee 
We will assume in what follows that the involved matrices are invertible square
matrices, unless otherwise noted.  However, in cases with
non-invertible $m_D$ (Sections \ref{sec:ss} and \ref{sec:0}) 
we will be able to exactly solve the problem
without the need of an expansion.   
If the eigenvalues of $M_R$ are all much heavier than the entries
of $m_D$, then this situation will be described as 
``$M_R \gg m_D$''. Block diagonalization of 
${\cal M}$ is now possible, and 
the block-diagonal matrices are approximately given by 
\be
- m_D^T \, M_R^{-1} \, m_D\quad \mbox{and}\quad 
M_R \, .
\ee
In Ref.~\cite{GL} a formalism to evaluate the corrections to these 
expressions to arbitrary order has been developed. 
Let us shortly summarize the derivation of that result. 
A unitary transformation diagonalizes ${\cal M}$ according to 
\be \label{eq:UM}
{\cal U}^T \, {\cal M} \, {\cal U} = 
\left( 
\baz
\tilde{m}_\nu & 0 \\
0 & \tilde{M}_R
\ea
\right)
\ee 
and transforms the states $(\nu_L, N_R^c)$ to the mass states 
$(\nu_{\rm l}, \nu_{\rm h})$, where the subscript ``l'' denotes 
light and ``h'' denotes heavy: 
\be 
{\cal U}^\dagger 
\left( 
\ba
\nu_L \\
N_R^c 
\ea
\right) 
= \left( 
\ba
\nu_{\rm l} \\
\nu_{\rm h}
\ea
\right)_L .
\ee
The matrix ${\cal U}$ can be written as \cite{GL} 
\be \label{eq:U}
{\cal U} = 
\left( \baz 
\sqrt{1 - B  B^\dagger} & B \\
-B^\dagger & \sqrt{1 - B^\dagger  B}
\ea
\right),\quad {\cal U}^\dagger = 
\left( \baz 
\sqrt{1 - B  B^\dagger} & -B \\
B^\dagger & \sqrt{1 - B^\dagger  B}
\ea
\right),
\ee
where $B$ is a complex $3 \times 3$ matrix (in general it has the
dimension of $m_D$), and the square root is
to be understood as 
\be \label{eq:series}
\sqrt{1 - B  B^\dagger} = 1 - \frac 12 \,  B  B^\dagger - 
\frac 18 \, B  B^\dagger \, B  B^\dagger - \ldots  
- \frac {\Gamma(-\frac 12 +n)}{n! \,\Gamma(-\frac 12)}\, (B  B^\dagger)^n - \ldots 
\ee

With this Ansatz the matrix ${\cal U}$ is unitary order by order in 
$B  B^\dagger$. The analogy of the form of ${\cal U}$ with a real 
two-by-two mixing matrix is obvious. We can insert ${\cal U}$ 
in Eq.~(\ref{eq:UM}) and the result for the three independent entries
of the r.h.s.~is 
\begin{align} \label{eq:12} \nonumber 
\sqrt{1 - B^\ast  B^T} \, m_D^T \, \sqrt{1 - B^\dagger  B} 
- B^\ast \, m_D \, B - B^\ast \, M_R \, \sqrt{1 - B^\dagger  B} & =  0
\, ,\\
-B^\ast \, m_D \, \sqrt{1-B B^\dagger} - \sqrt{1-B^\ast B^T} \, m_D^T
\, B^\dagger + B^\ast \, M_R \, B^\dagger & =  \tilde{m}_\nu  \, , \\
\sqrt{1-B^T B^\ast} \, m_D \, B + B^T \, m_D^T \, \sqrt{1-B^\dagger B}
+\sqrt{1-B^T B^\ast} \, M_R \, \sqrt{1-B^\dagger B} & =  \tilde{M}_R \nonumber 
\, . 
\end{align}
Now the see-saw approximation enters the game, by assuming that 
$B$ can be written as a power series in terms of $1/M_R$, i.e., in
terms of the eigenvalues of $M_R$, which are assumed to be much
heavier than the entries of $m_D$. Hence, 
$B = B_1 + B_2 + \ldots$, where $B_i$ is of order $(1/M_R)^i$. The
square root then reads 
\be
\sqrt{1 - B^\dagger  B} \simeq 1 - \frac 12 \, B_1^\dagger B_1 -
\frac12 
\left(B_1^\dagger B_2 + B_2^\dagger B_1  \right) - \ldots 
\ee
A recursive solution of Eq.~(\ref{eq:12}) is now possible. At leading
order the solution of  Eq.~(\ref{eq:12}) is given 
by 
\be \label{eq:B1}
B_1^\ast = m_D^T \, M_R^{-1} \,. 
\ee
The next order term $B_2$ in 
the expansion is, in the limit of a vanishing triplet contribution,
zero. This is true for all $B_i$, where $i$ is even \cite{GL}. We
obtain for the 3rd and 5th order terms 
\begin{align} \nonumber 
B_3^\ast \, M_R =& 
-\frac 12 \, B_1^\ast \, B_1^T \, m_D^T - \frac 12 \, m_D^T \,
B_1^\dagger \, B_1 - B_1^\ast \, m_D \, B_1 + \frac 12 \, 
B_1^\ast \, M_R \, B_1^\dagger \, B_1 \\ 
 = &-\frac 12 \, m_D^T \, M_R^{-1} \, (M_R^\ast)^{-1} \, m_D^\ast \,
m_D^T - m_D^ T \, M_R^{-1} \, m_D \, m_D^\dagger \, (M_R^\ast)^{-1} \,
, \\
B_5^\ast \, M_R = &-\frac 12 \,\left(B_1^\ast  \, B_3^T+B_3^\ast  \,
B_1^T + \frac 14  \, B_1^\ast  \, B_1^T  \, B_1^\ast  \, B_1^T\right) 
m_D^T -B_1^\ast  \, m_D  \, B_3 \nonumber\\
& - B_3^\ast  \, m_D  \, B_1  + \frac 12  \, B_3^\ast  \, M_R  \,
B_1^\dagger  \, B_1 \, . \nonumber 
\end{align}
Inserting $B_1$ and $B_3$ in the 11- and 22-entries of 
Eq.~(\ref{eq:UM}) yields 
\begin{align}  \nonumber 
\tilde{m}_\nu = & 
-  m_D^T \, M_R^{-1} \, m_D + \frac 12 \, m_D^T \, M_R^{-1} 
\left[ 
m_D \, m_D^\dagger \, (M_R^\ast)^{-1} + (M_R^\ast)^{-1} \, m_D^\ast
\, m_D^T 
\right] M_R^{-1} \, m_D \, , \\ 
\tilde{M}_R = & M_R + \frac 12 \left[ 
m_D \, m_D^\dagger \, (M_R^\ast)^{-1} + (M_R^\ast)^{-1} \, m_D^\ast
\, m_D^T 
\right] \, .\nonumber
\end{align}
One is lead to define the symmetric matrix 
\be\label{eqn:XandA}
X \equiv A + A^T \,, \mbox{ where } A \equiv 
m_D \, m_D^\dagger \, (M_R^\ast)^{-1} \, . 
\ee
The order of magnitude of $X$ is $m_D^2/M_R$ and 
one can simplify the relations to 
\begin{subequations}\label{eqn:mnuMR}
\begin{eqnarray}
\tilde{m}_\nu &=& 
- \, m_D^T \, M_R^{-1} \, m_D +\frac 12 \, m_D^T \, M_R^{-1} 
\, X \, M_R^{-1} \, m_D \, ,\label{eqn:mnu} \\
\tilde{M}_R &=& M_R + \frac 12 \, X \, .\label{eqn:MR}
\end{eqnarray}
\end{subequations}
For completeness, we also give the NNLO terms to $m_\nu$ and $M_R$,
which are 
\begin{align} \nonumber
 \tilde{m}_\nu^{\rm NNLO} = & 
\frac 12 \, m_D^T \, M_R^{-1} \, \bigg[ 
\frac 14 \, A \, M_R^{-1} \, A + \frac 14 \, A^T \, M_R^{-1} \, A^T +
\frac 12 \, A^T \, M_R^{-1} \, A 
+ \frac 12 \, (M^\ast_R)^{-1} \, A^\ast \, A^T \nonumber\\
& \left. + \frac 12 \, A \, A^\dagger \, (M^\ast_R)^{-1} + A \, A^\ast
\, (M^\ast_R)^{-1} + (M^\ast_R)^{-1} \, A^\dagger \, A^T \right]  \,
M_R^{-1} \, m_D\nonumber \, , \\  \nonumber
\tilde{M}_R^{\rm NNLO} = & 
-\frac 12 \, \bigg[ 
\left.  A \, A^\ast \, (M^\ast_R)^{-1} + (M^\ast_R)^{-1} \, A^\dagger
\, A^T + \frac 14 \, A \, M_R^{-1} \, A + \frac 14 \, A^T \, M_R^{-1} \,
A^T \right] . 
\label{eqn:NNLO}
\end{align}\noindent
The zeroth order terms of the light and heavy mass matrices
are $m_D^2/M_R$ and $M_R$, respectively. The relative NLO corrections 
are of order 
$X /M_R = m_D^2/M_R^2$ for both. The absolute correction is of order 
$m_D^4 /M_R^3$ for the light neutrinos and $m_D^2/M_R$ for the heavy
neutrinos. Note that this NLO correction vanishes when $A$ is
antisymmetric. 
The absolute order of magnitude of the NNLO terms is $m_D^6 /M_R^5$
for the light neutrinos and $m_D^4 /M_R^3$ for the heavy ones. In
general, the N$^{n+1}$LO term of the heavy neutrinos has the same 
absolute order of magnitude than the N$^n$LO term of the light ones. \\

A comment to be made here is that the same expressions for the
corrections are obtained in type III see-saw scenarios \cite{seesaw3},
for which $M_R$ is the mass term of the neutral component of a weak
fermion triplet. 

In this work we will mostly ignore the possibility of the presence of
a Higgs triplet, which would fill the upper left entry of ${\cal M}$
in Eq.~(\ref{eqn:typeIMass}) with a term $m_L$ \cite{seesaw2}. 
In this case, the first order correction to $m_\nu$ is \cite{GL} 
\begin{eqnarray}
\tilde{m}_\nu &=& m_L - 
m_D^T \, M_R^{-1} \, m_D +\frac 12 \, m_D^T \, M_R^{-1} 
\, X \, M_R^{-1} \, m_D  - \frac 12 \, (C + C^T) 
\, , \label{eq:mlcor}
\end{eqnarray}
where $C = m_D^T \, M_R^{-1}  \, (M_R^\ast)^{-1} \, m_D^\ast \,
m_L$.

\subsection{\label{sec:app}Some possible Applications}

We will continue with a few examples on the possible consequences of the NLO
terms. 

The typical order of magnitude of the terms 
is $m_D \simeq 10^2$ GeV and $M_R
\simeq 10^{14}$ GeV, for which $m_\nu \simeq 0.1$ eV. In this case, of
course, the NLO terms are negligible (the same is true for the
unitarity violation, see below). 
An exception is when the
Majorana singlets are put to TeV scale, which is largely 
motivated by current collider opportunities. The mixing with the
singlets is naively of order $m_D/M_R$, and hence the requirement of 
$m_\nu = 0.1$ eV would lead to small $m_D$ and thus small
mixing. However, it is possible that small neutrino masses are an
effect of cancellation of large terms, with $M_R
\simeq$ TeV and $m_D \simeq v$, in which case the ratio of leading order 
and NLO terms is $m_D^2/M_R^2 \sim 10^{-2}$, a percent effect! 
In the light of future precision experiments, this is a correction 
one surely should take into account. 

Let us give an illustrative example on this. We enter now the basis in
which $M_R$ is real and diagonal. For a two neutrino case, and a diagonal
right-handed neutrino mass matrix 
$M_R$, and if the Dirac mass matrix is written in its most general form 
\be
m_D = \left(\baz 
a & b \\
c & d 
\ea
\right),
\ee
the neutrino mass matrix is at leading order: 
\be
m_\nu = \frac{1}{M_1} \left( \baz 
a^2 & a\, b\\
\cdot & b^2  
\ea
\right) + 
\frac{1}{M_2} \left( \baz 
c^2 & c\, d\\
\cdot & d^2  
\ea
\right) 
\, . 
\ee
One might imagine that $M_{1,2}$ lie around TeV, 
$m_D$ lies around $v$, and that at leading
order the neutrino mass matrix vanishes, $m_\nu = 0$. 
In order for $m_\nu $ to vanish\footnote{This will hold true 
to all orders, see below.}, the requirements
\be
m_D = \left( 
\baz 
x & a  x \\
y & a  y 
\ea
\right) \mbox{ and }~
\frac{x^2}{y^2} = - \frac{M_1}{M_2} 
\ee
must hold simultaneously \cite{zero}. A small correction in one or
both of those two conditions will generate small but non-zero neutrino
mass. For instance, violating the second condition as 
$x = i y \, \sqrt{\frac{M_1}{M_2}} \, (1 + \epsilon)$, one finds 
\be
m_\nu^0 = -m_D^T \, M_R^{-1} \, m_D 
= \frac{y^2}{M_2} \, \epsilon \, (2 + \epsilon) 
\left( \bad 
1 & a \\
\cdot & a^2 
\ea
\right) , 
\ee
which has eigenvalues $m_1 = 0$ and $m_2 = (1 + a^2) \,
y^2/M_2 \, (2 + \epsilon) \, \epsilon$, obtained by diagonalizing
$m_\nu^0$ with the mixing matrix 
\[ 
U^0 = \frac{1}{\sqrt{1 + a^2}} \left( \baz 
-a & 1 \\
1 & a 
\ea
\right) . 
\]  
Note that large mixing would imply $a = {\cal O}(1)$. 
For, $a = 1$, $y \simeq 100$ GeV and $M_2 \simeq $ TeV, it follows
that $\epsilon \simeq 10^{-12}$ to give $m_2 = 0.05$ eV. This
illustrates the enormous tuning which has to present in order for this
mechanism to work. 

What are the corrections to $m_\nu^0$? We have shown above that the
NLO term to $m_\nu^0$ is $\frac 12 m_D^T \, M_R^{-1} \, X \, M_R^{-1}
\, m_D$, where $X = A + A^T$ and $A = m_D \, m_D^\dagger \,
(M_R^{-1})^\ast$. Evaluating this with our example gives 
\be
m_\nu^1 = -\frac{(2 + \epsilon) \, \epsilon}{M_1 \, M_2^3} 
\left( 
M_1 + (1 + \epsilon)^2 M_2 
\right) y^4 \, (1 + a^2) 
\left( \baz 
1 & a \\
\cdot & a^2
\ea
\right) . 
\ee
Comparing the zeroth and first order term, we have 
$m_\nu^0 = {\cal O} (\epsilon \, y^2/M)$ and $m_\nu^1 = {\cal O} 
(\epsilon \, y^4/M^3)$, which for $y \simeq 100$ GeV and 
$M \simeq $ TeV results in a NLO term being suppressed only at the 
percent level. The mixing matrix
stays the same in this example, and the non-zero eigenvalue of  
$m_\nu^0 + m_\nu^1$ is different from the non-zero eigenvalue of
$m_\nu^0$ by order $\epsilon \, y^4/M^3$. It should be clear to
realize that the moderate hierarchy between the submatrices 
$m_D$ and $M_R$ of ${\cal M}$ leads here to sizable effects.\\

One may wonder what happens when $m_\nu = 0$ at leading
order, for instance if
\be
m_D = \left( 
\bad 
0 & 0 & 0 \\
0 & 0 & 0 \\
a_3 & b_3 & c_3 
\ea 
\right)~\mbox{ and }~
M_R = \left( 
\bad 
0 & 0 & M_1 \\
\cdot & M_2 & 0 \\
\cdot & \cdot & 0 
\ea 
\right) . 
\ee
However, one can show \cite{KPL,GL} that the vanishing of $m_\nu$ 
will remain true to all orders\footnote{A proof for the case when a
triplet is present can be found in \cite{xing}.} (the rank of ${\cal M}$ is three). 
 We will discuss the
realistic case of one vanishing eigenvalue of $m_\nu$ in Section
\ref{sec:0}.\\

Another aspect is that zeros entries could be filled by NLO terms. 
Consider the case 
\be
m_D = \left( 
\bad 
0 & 0 & c_1 \\
0 & b_2 & c_2 \\
a_3 & b_3 & 0   
\ea 
\right)~\mbox{ and }~
M_R = \left( 
\bad 
M_1 & 0 & 0 \\
\cdot & 0 & M_2 \\
\cdot & \cdot & 0 
\ea 
\right) .
\ee
The resulting low energy mass matrix at zeroth order is 
\be
m_D^T \, M_R^{-1} \, m_D = \left( 
\bad 
0  & \frac{a_3 \, b_2 }{M_2} & \frac{a_3 \, c_2 }{M_2} \\
\cdot & 2\frac{b_2 \, b_3}{M_2} & \frac{b_3 \, c_2}{M_2} \\
\cdot & \cdot & \frac{c_1^2}{M_3}
\ea
\right) 
\ee
and forbids neutrino-less double beta decay because the 11-entry of
$m_\nu$ vanishes. The NLO term fills this entry with a contribution 
\be
(m_\nu)_{11} = \frac{a_3^2 \, b_2 \, b_3}
{M_2^3 }  \, .
\ee
However, this term is suppressed with respect to the non-zero entries
in $m_D^T \, M_R^{-1} \, m_D$ by many orders of magnitude, and can
safely be neglected. It is typically even smaller than terms of order 
$U_{e i}^2 \, m_i^3/q^2$ ($q^2 \simeq 0.1$ GeV$^2$ the typical
momentum exchange in double beta decay), which are in general non-zero
when $U_{e i}^2 \, m_i = 0$. 
An exception could be
when such zero textures are generated in scenarios in which small
$m_\nu$ is generated by cancellations of large terms, see above.

\subsection{Connection to Unitarity Violation}\label{sec:UnitarityViolation}

It is well-known that an intrinsic unitarity violation is 
present in those see-saw scenarios which contain extra fermions 
(i.e., not in a pure type II see-saw). This can be shown easily by
diagonalizing Eq.~(\ref{eq:UM}) in a slightly different way, namely
via 
\be \label{eq:Uneu}
{\cal U} = \left( \baz
N & S \\
T & V 
\ea \right) , 
\ee
where $N,S,T,V$ are $3\times3$ mixing matrices which are in general 
non-unitary. By evaluating 
\be
{\cal U}^T \, {\cal M}_\nu \,  {\cal U} = 
 \left( \baz 
m_\nu^{\rm diag} & 0 \\
0 & M_R^{\rm diag}
\ea \right) .
\ee
and by assuming that $S,T$ are of order $m_D/M_R$, one obtains from
the 12-entry of ${\cal U}^T \, {\cal M}_\nu \,  {\cal U}$ that at
leading order $T^T \simeq -N^T \, m_D^T \, M_R^{-1}$. Inserting this
in the 11-entry gives 
\be
m_\nu^{\rm diag} \simeq -N^T \, m_D^T \, M_R^{-1} \, m_D \, N \,  
\ee
and $V^T \, M_R \, V \simeq M_R^{\rm diag}$. The PMNS matrix $N$ is
therefore non-unitary, because $N N^\dagger = \mathbbm{1} - S 
S^\dagger$ and $N^\dagger N = \mathbbm{1} - T^\dagger
T$. Phenomenologically, the non-unitarity can be described by writing 
\be \label{eq:N}
N = (\mathbbm{1} + \eta) \, U_0 \, , 
\ee
where $U_0$ is unitary and $\eta$ hermitian. The latter matrix
contains three phases, but current constraints exist only for its
absolute values \cite{NU}: 
\be \label{eq:etalim}
|\eta| < \left( \bad
4.0 \times 10^{-3} & 1.2 \times 10^{-5} & 3.2 \times 10^{-3} \\
\cdot & 1.6 \times 10^{-3} & 2.1 \times 10^{-3} \\ 
\cdot & \cdot & 5.3 \times 10^{-3} 
\ea \right) . 
\ee
By comparing Eq.~(\ref{eq:N}) with (\ref{eq:Uneu}) we can identify 
$\eta \simeq -\frac 12 \, S S^\dagger$. By inserting in 
$N^\dagger N = \mathbbm{1} - T^\dagger T$ the above relation for $T$
and the phenomenological description for $N$ from Eq.~(\ref{eq:N}),
one finds 
\be \label{eq:etaB}
\eta \simeq -\frac 12 \, m_D^\dagger \left( M_R^{-1} \right)^\ast \, 
M_R^{-1} \, m_D = -\frac 12 \, B_1  B_1^\dagger \, . 
\ee 
Hence, we can read off the amount of unitarity
violation from the first
order expression of $B$, which is given as 
$B_1 = m_D^\dagger \, (M_R^\ast)^{-1}$ in
Eq.~(\ref{eq:B1})\footnote{Actually, the above limits on $\eta$ 
often assume that the physics leading to unitarity violation is
inaccessible at low energy. For instance, the amplitudes of lepton
flavor violating charged lepton transition $\ell_i \to \ell_j$ receive
contributions from fermion singlets in the form of 
$\sum_k {\cal U}_{jk} \, {\cal U^\dagger}_{ki} \, g(x_k)$, where 
$x_k = m_k^2/m_W^2 $ in the loop function $g(x)$, and $m_k$ is a
singlet fermion which mixes with the SM particles. For singlet masses
above a few 100 GeV the limits are basically equivalent to (\ref{eq:etalim}).}. 
If a triplet term $m_L$ is present, the same calculation can be
performed and the result from Eq.~(\ref{eq:etaB}) stays the same as long as $m_L \ll M_R$. 
A triplet term does therefore not induce unitarity violation. 
As we will see in Section \ref{sec:var}, we can calculate the NLO
terms for see-saw variants like the inverse or double see-saw in the
same way as we have done above for the type I
see-saw. This makes it
possible to simply write down the magnitude and structure 
of unitarity violation for those scenarios.

\section{\label{sec:exa}Special Cases in Type I See-Saw} 
There are certain cases in which the specific (flavor) structure of the
mass matrices leaves imprints on the higher order see-saw
corrections and the unitarity violating parameters. We will discuss
some examples, starting first with $\mu$--$\tau$ symmetry and then 
move on to scaling ($m_3 = \theta_{13} = 0$ in the inverted
hierarchy), which we will generalize to scenarios containing a vanishing
neutrino mass with arbitrary mixing and mass ordering.  

\subsection{$\mu$--$\tau$ Symmetric See-Saw}
Consider $\mu$--$\tau$ symmetric\footnote{Actually, this is a
  generalized form of the usually considered $\mu$--$\tau$ symmetry,
  which denotes the invariance under exchange of flavor indices $\mu$
  and $\tau$ in $m_\nu$. A more correct name would be 2--3 symmetry,
  but we stick to the name $\mu$--$\tau$ symmetry.} 
$m_D$ and $M_R$, 
i.e., \cite{mtss} 
\be \label{eq:mtss}
M_R = \left( \bad 
X & Y & Y \\
\cdot & W & Z \\
\cdot & \cdot & W
\ea
\right) ~\mbox{ and } 
m_D = \left( \bad 
a & b & b \\
d & e & f \\
d & f & e 
\ea \right) . 
\ee
As a result of such a structure the leading term of the low energy
mass matrix is 
\be
m_\nu^0 = - m_D^T \, M_R^{-1} \, m_D = 
\left( \bad 
A & B & B \\
\cdot & D & E \\
\cdot & \cdot & D 
\ea \right) , 
\ee
where $A,B,D,E$ are functions of the parameters in $m_D$ and $M_R$. 
The above matrix $m_\nu^0$ predicts to the eigenvalue 
$D - E$ the eigenvector $(0,-1,1)^T$. 
Hence, if $D - E $ corresponds to the largest (smallest) mass, 
$\theta_{13} = 0$ and $\theta_{23} = \pi/4$ in the normal (inverted)
mass ordering is predicted. Interestingly, the unitarity violating
parameter $\eta$ is also $\mu$--$\tau$ symmetric, 
\be
\eta = -\frac 12 \, m_D^\dagger \left( M_R^{-1} \right)^\ast \, 
M_R^{-1} \, m_D = 
\left( \bad 
|x| & y & y \\
y^\ast & |z| & w \\
y^\ast & w^\ast & |z| 
\ea \right) , 
\ee
where the new parameters $x,y,z,w$ are functions of the entries in 
$m_D$ and $M_R$ in Eq.~(\ref{eq:mtss}). 
This implies in particular that $\eta_{e\tau}$ is predicted to be 
extremely small and below values which can be probed in future
neutrino oscillation facilities \cite{NU_fut}. 

The eigenvalue $D - E = (e-f)^2/(w-z)$ of the zeroth order matrix is
in its exact form (that is, by diagonalizing ${\cal M}$ instead of 
$m_\nu^0$) given as 
\be
m_3 = \frac 12 \left( 
(z - w) - \sqrt{(z-w)^2 + 4 \, (e-f)^2}
\right) \equiv 
M - \sqrt{M^2 + 4 \, m^2} \, . 
\ee
The exact eigenvector to this eigenvalue can be written as 
\be
 \left( 
\ba
U_{e3} \\
U_{\mu 3} \\
U_{\tau 3} \\
U_{{N_1}3} \\
U_{{N_2}3} \\
U_{{N_3}3} 
\ea \right) = N 
\left( 
\ba
0 \\
-1 \\
1 \\
0 \\
-\frac{1}{2 m} \left(M - \sqrt{M^2 + 4 \, m^2} \right) \\
\frac{1}{2 m} \left(M - \sqrt{M^2 + 4 \, m^2} \right) 
\ea
\right) , 
\ee
where we included its normalization in the factor $N$. We therefore 
showed that 
$U_{e3} = 0$ and maximal mixing in the sense $|U_{\mu 3 }| = |U_{\tau
3}|$ is not modified by higher order corrections in 
$\mu$--$\tau$ symmetric see-saw scenarios. Interestingly, the state
with mass $m_3$ does not mix with one heavy neutrino and mixes in equal
amounts with the other two.

\subsection{\label{sec:ss}Scaling}

The next example deals with ``scaling'' \cite{scal0} (see also \cite{others}), for which 
the Dirac mass matrix has the following texture: 
\be  \label{eq:scal}
m_D = \left( \bad 
a_1 & b_1 & b_1/c \\
a_2 & b_2 & b_2/c \\
a_3 & b_3 & b_3/c  
\ea \right) . 
\ee 
The third column is proportional to the second one, the relevant
factor being the ``scaling constant'' $c$. Interestingly, independent
on the form of $M_R$ (other than being non-singular) the low energy mass
matrix has the form \cite{JR}
\be
m_D^T \, M_R^{-1} \, m_D = \left( 
\bad 
A & B & B/c \\
\cdot & D & D/c \\
\cdot & \cdot & D/c^2 
\ea
\right) . 
\ee
Such a low energy mass matrix has been derived for instance in
explicit flavor symmetry models based on $D_4$ in Ref.~(\cite{scal0}).
The prediction of this particular texture is that the eigenvector to
the zero eigenvalue (note that the rank of $m_D^T \, M_R^{-1} \, m_D$
is 2) is $(0,-1/c,1)^T$, and hence scaling predicts an inverted
hierarchy, with $U_{e3} = 0$ and $\tan^2 \theta_{23} = 1/|c|^2$ \cite{scal0}. 

It is easy to see that with $m_D$ given in Eq.~(\ref{eq:scal}) 
the full $6\times6$ mass matrix ${\cal M}$ has
rank 5 and the eigenvector corresponding to the zero eigenvalue is 
\be
\left( 
\ba
U_{e3} \\
U_{\mu 3} \\
U_{\tau 3} \\
U_{{N_1}3} \\
U_{{N_2}3} \\
U_{{N_3}3} 
\ea \right) = \frac{1}{\sqrt{1 + |1/c|^2}} 
\left( 
\ba
0 \\
-1/c \\
1 \\
0 \\
0 \\
0
\ea
\right) . 
\ee
Therefore, there are no corrections to the predictions $U_{e3} = 0$
and $\tan^2 \theta_{23} = 1/|c|^2$ in see-saw scenarios obeying
scaling. The massless neutrino does not mix with the 
heavy ones. The unitarity violation obeys the relations 
\be
\left|\frac{\eta_{e\mu}}{\eta_{e\tau}}\right| = 
\left|\frac{\eta_{\mu\mu}}{\eta_{\mu\tau}}\right| = 
\left|\frac{\eta_{\tau\mu}}{\eta_{\tau\tau}}\right| = |c| = \cot \theta_{23}
\, , 
\ee
and again the implied value of $\eta_{e\tau}$ is very small. 

\subsection{\label{sec:0}Vanishing Eigenvalue}
The specific example discussed in the last subsection had a vanishing neutrino
mass in $m_\nu^0 = -m_D^T \, M_R^{-1} \, m_D$, and higher order
corrections did not induce a non-zero mass (this is actually trivial,
since the rank of ${\cal M}$ is five), nor did they modify the
mixing matrix elements of the vanishing eigenvalue. 
Is this true in general? In what follows we will show that this is indeed the case. 

If $m_\nu^0$ is to have rank 2, it follows that there is an
eigenvector $|{\boldsymbol \psi} \rangle$ to it such that 
$m_\nu^0 \, |{\boldsymbol \psi} \rangle =
0$. If $M_R$ is non-singular (we treat this case of det$\,M_R=0$ later)
then this means that $m_D$ has rank 2. Hence, there is an eigenvector 
$|\Phi \rangle$ to $m_D$ with the property $m_D \, |\Phi \rangle  =
0$. With the definition of $m_\nu^0$ it follows that $m_\nu^0 \, |\Phi 
\rangle = 0$. Since $m_\nu^0$ can not have two zero eigenvalues, 
$|\Phi \rangle$ must be proportional to $|{\boldsymbol \psi} \rangle$, and hence 
$m_D \, |{\boldsymbol \psi} \rangle = 0$, or to be more specific: 
\be \label{eq:md310}
m_D  \left( \ba 
U_{e1}  \\
U_{\mu 1} \\
U_{\tau 1} 
\ea \right) = 0~\mbox{ or } 
m_D  \left( \ba 
U_{e3}  \\
U_{\mu 3} \\
U_{\tau 3} 
\ea \right) = 0 \, , 
\ee
depending on whether the normal or inverted ordering is present. Note
again that this is independent on the form of $M_R$. Taking the normal
ordering as an example, Eq.~(\ref{eq:md310}) implies that the Dirac mass 
matrix takes the form
\be \label{eq:mdrank2} 
m_D = \left( \bad 
a_1 & b_1 & - \frac{U_{e1} \, a_1 + U_{\mu 1} \, b_1}{U_{\tau 1}}\\
a_2 & b_2 & - \frac{U_{e1} \, a_2 + U_{\mu 1} \, b_2}{U_{\tau 1}}\\
a_3 & b_3 & - \frac{U_{e1} \, a_3 + U_{\mu 1} \, b_3}{U_{\tau 1}}
\ea \right) ,
\ee
or similar relations in the first or second column of $m_D$. 
In case of an inverted hierarchy, $U_{\alpha 1}$ has to be replaced
with $U_{\alpha 3}$, and if in this case $U_{e3} = 0$, we have the
scaling scenario described above in Section \ref{sec:ss}. 
Inserting Eq.~(\ref{eq:mdrank2}) in the $6\times6$ matrix ${\cal M}$ reveals that
it has rank 5, and the exact eigenvector to the zero mass state is
simply 
\be \label{eq:bb}
\left( 
\ba
U_{e i} \\
U_{\mu i} \\
U_{\tau i} \\
0 \\
0 \\
0
\ea
\right) , 
\ee
with $i = 1$ (3) for the normal (inverted) ordering. Thus, if there is
a vanishing eigenvalue $m_i$ of $m_\nu^0$, then there are no corrections
arising to its mixing parameters $U_{\alpha i}$, where $\alpha = e, 
\mu, \tau$. In addition, this neutrino does not mix with heavy ones.\\

Is this conclusion valid in both ways? Let us assume that 
the $6\times6$ mass matrix ${\cal M}$ has a vanishing eigenvalue,
i.e., 
\be
{\cal M} \, \vec a = 0 \, , 
\ee
where $\vec a = (a_1, a_2, a_3, a_4, a_5, a_6)^T$ is the eigenvector of the
zero eigenvalue. Solving this equation for, say, the third column of
$m_D$ and $a_{4,5,6}$ gives nothing but Eqs.~(\ref{eq:mdrank2}) 
and (\ref{eq:bb}).  Hence, we have shown that
the eigenvector to a vanishing neutrino mass receives no corrections 
from higher order see-saw terms.

\subsection{\label{zero}Almost vanishing $m_\nu$}
Let us return to the scenarios in which small $m_\nu$ is generated by
a small perturbation to $m_\nu=0$. In a 3 family 
framework the condition for vanishing $m_\nu$ is for diagonal
$M_R$ that \cite{zero}
\be \label{eq:zero3}
m_D = \left( 
\bad 
x & a \, x & b \, x \\ 
y & a \, y & b \, y \\ 
z & a \, z & b \, z 
\ea \right)~\mbox{ and }~
\frac{x^2}{M_1} + \frac{y^2}{M_2} + \frac{z^2}{M_3} = 0 \, .
\ee
The generalization to non-diagonal $M_R$ has recently been discussed
in Ref.~\cite{inder}, and the possibility of percent effects of the NLO
terms has been discussed in Section \ref{sec:app}. 
The structure of the implied unitarity violation has
been analyzed recently in Ref.~\cite{HZ}. With Eq.~(\ref{eq:zero3})
and the definition of $\eta$ in terms of $B_1$ we have 
\be
\eta = -\frac 12 \left( 
\frac{x^2}{M_1} +  \frac{y^2}{M_2} + \frac{z^2}{M_3} 
\right) 
\left( 
\bad 
1 & a & b \\
a^\ast & |a^2| & a^\ast \, b \\
b^\ast & a \, b^\ast & |b|^2
\ea 
\right) . 
\ee
As $|\eta_{e\mu}|$ is known to be very small, $a$ can essentially be set to zero
and the flavor structure of $\eta$ becomes very simple \cite{HZ}.

\section{\label{sec:var}NLO Terms to See-Saw Variants}
There are popular variants of the type I see-saw, in which additional 
singlets $S$ are added to the theory, and the 
$(\nu_L^c, N_R)$ basis is extended to a $(\nu_L^c, N_R, S)$ basis: 
\be \label{eq:master} 
{\cal M} = 
\left( 
\bad
m_L & m_D^T & m_{DS}^T \\
m_D & M_R & m_{RS}^T \\
m_{DS} & m_{RS} & M_S \ea 
\right) .
\ee
The diagonal entries are complex symmetric, while the off-diagonal
elements are arbitrary complex matrices. 
As mentioned above, we will not consider the presence of a triplet
term $m_L$ here. The frequently discussed variants of the type 
I see-saw are obtained from 
this equation by setting some terms to zero and assuming a hierarchy
in the eigenvalues of the surviving terms. We will discuss in the 
following these variants and apply the formalism discussed in 
Section \ref{sec:GL} to analyze the NLO terms and the order of magnitude 
of the unitarity violation.  
The results of this Section are summarized in Table \ref{tab:tab}.

\subsection{\label{sec:double}Double See-Saw} 
In the double see-saw scenario we have \cite{double}: 
 \be \label{eq:exa1} 
{\cal M} = 
\left( 
\bad
0 & m_D^T & 0 \\
m_D & 0 & m_{RS}^T \\
0 & m_{RS} & M_S \ea 
\right)
\ee
with the conditions $m_D,m_{RS} \ll M_{S}$ and $m_D \ll m_{RS}^2/M_S$.
To block-diagonalize ${\cal M}$, we define
\begin{equation}\label{eqn:mathbbMDMR}
\mathbb M_D:=
\left(\ba
m_D\\
0
\ea\right)\qquad \mathrm{and} \qquad \mathbb M_R:=\left(\baz
0&m^T_{RS}\\
m_{RS}&M_S\ea\right) ,
\ee
and write
\be\label{eqn:DoubleMass}
\cal M = \left(
\baz 
0 & \mathbb M_D^T\\
\mathbb M_D & \mathbb M_R
\ea\right) . 
\ee
The eigenvalues of the symmetric block $\mathbb M_R$ are of order 
$M_S$ and $m_{RS}^2/M_S$ which are, because of the above mentioned conditions, 
much bigger than the entries in $\mathbb M_D$. 
Thus, if we compare Eq.~\eqref{eqn:DoubleMass} with 
Eq.~\eqref{eqn:typeIMass}, we recognize that the double see-saw formulas and 
their corrections simply 
follow from the type I equations which we presented in Section 
\ref{sec:GL}. We only have to perform the substitutions
\be\label{eqn:Subst}
 m_D \rightarrow \mathbb M_D\;, \qquad M_R \rightarrow \mathbb M_R\;.
\ee
Note that $\mathbb M_R$ has the same structure as the type I 
see-saw matrix (\ref{eqn:typeIMass}) whose inverse is given in 
Eq.~(\ref{eq:Mi}), and the inverse of $\mathbb M_R$ can therefore simply 
be read off from that expression. 
For illustration, let us determine the usual double see-saw formula and 
its first order correction. 
The relevant relations are given in Eq.~\eqref{eqn:mnuMR}. 
Applying \eqref{eqn:Subst}, equation \eqref{eqn:XandA} translates into 
\be\label{Xdouble}
X = A + A^T \,, \mbox{ where } A = 
\mathbb M_D \, \mathbb M_D^\dagger \, (\mathbb M_R^\ast)^{-1} \; ,
\ee
and we obtain for the mass of the lightest neutrinos the expression
\be\label{eqn:lightdouble}
\tilde m_\nu = m_D^T \, m_{RS}^{-1} \, M_S \, 
(m_{RS}^T)^{-1} \, m_D + \mathcal O\left( M_S 
\frac{m_D^4 }{m_{RS}^4}  \, 
\left(1 +  \frac{M_S^2}{m_{RS}^2} \right) \right) ,
\ee
which follows from Eq.~\eqref{eqn:mnuMR} after inserting 
Eq.~(\ref{eqn:Subst}). 
The first term on the right-hand side represents the well known 
double see-saw formula. By setting in the suggestive values $m_D
\simeq 100$ GeV, $M_S \simeq M_{\rm Pl}$ and 
$m_{RS} \simeq M_{\rm GUT} \simeq 10^{16}$ GeV, we can generate the
correct order of magnitude for neutrino masses. 
The second term in Eq.~(\ref{eqn:lightdouble}) gives the order of 
magnitude of the very lengthy NLO corrections, which however can be 
obtained in a straightforward manner by inserting Eq.~(\ref{Xdouble}) in 
Eq.~(\ref{eqn:mnuMR}). For the sake of completeness, let us quote the
result: 
\bea \label{eq:double}
-2 \, m_\nu^1 = \\
 m_D^T \, m_{RS}^{-1} \, M_S \, (m_{RS}^T)^{-1} \left( 
m_D \, m_D^\dagger \, (m_{RS}^\ast)^{-1} \, M_S^\ast \,
(m_{RS}^\dagger)^{-1} + \mbox{\small (last term)$^T$} \right) 
m_{RS}^{-1} \, M_S \, (m_{RS}^T)^{-1} \, m_D \\
+\, m_D^T \, m_{RS}^{-1}\left(M_S \, (m_{RS}^T)^{-1} \, m_D \, m_D^\dagger
\, (m_{RS}^\ast)^{-1} + \mbox{\small (last term)$^T$}\right) (m_{RS}^T)^{-1} \, m_D  \, . 
\eea 
The first term of order $M_S^3 \, m_D^4 \, m_{RS}^{-6}$ is the leading one
for the double see-saw. 
By setting in the suggestive values given above,  
we realize that the latter gives a contribution of the same order, if
not larger, than the correction $m_D^4/M_R^3$ for the type I see-saw 
formula. It is however, with the indicated values of 
$m_D$, $M_S$ and $m_{RS}$ a negligibly small contribution, 
which may change in other realizations. 

Let us now determine the unitarity violation. 
From Section \ref{sec:UnitarityViolation} we know that we can read off
its amount from the first order expression 
of $B$ (cf.~Eq.~\eqref{eq:etaB}). Thus,
\be
\eta \simeq 
-\frac 12 \, \mathbb M_D^\dagger \left( \mathbb M_R^{-1} \right)^\ast \, 
\mathbb M_R^{-1} \, \mathbb M_D\,,
\ee
which has the following explicit form: 
\bea\label{eqn:etadouble}
\eta \simeq -\frac 12 \left( 
m_D^\dagger \, (m_{RS}^\ast)^{-1} \, M_S^\ast \, (m_{RS}^\dagger)^{-1}
\, m_{RS}^{-1} \, M_S \, (m_{RS}^T)^{-1} \, m_D + 
m_D^\dagger \, (m_{RS}^\ast)^{-1} \, (m_{RS}^T)^{-1} \, m_D \right) \\
= {\cal O}
\left(\frac{m_D^2}{m_{RS}^2} \left(1 + \frac{M_S^2}{m_{RS}^2}\right) \right).
\eea
Again, for the double see-saw the term $M_S^2\,m_D^2/m_{RS}^4$ in the
second row, giving the
order of magnitude of $\eta$, is expected to be the dominating one.

Finally, we mention the possibility of ``screening'' \cite{screening},
in which case $m_D = \epsilon \, m_{RS}^T$. It follows that the zeroth
plus first order terms are given as 
\be
\tilde m_\nu = \epsilon^2 \, M_S - \frac 12 \, \epsilon^4 \, M_S
\left( 
M_S^\ast \, (m_{RS}^\dagger)^{-1} + (m_{RS}^\ast)^{-1} \, M_S^\ast 
\right) M_S - \epsilon^4 \, M_S\, . 
\ee
One sees that the leading order term has its flavor structure
determined by the high (possibly Planck) scale physics, while the
corrections include additional flavor terms. The unitarity
violation simplifies to 
\be
\eta = -\frac 12 \, \epsilon^2 \left( 
\mathbbm{1} + M_S^\ast \, (m_{RS}^\dagger)^{-1} \, m_{RS}^{-1} \, M_S \right) .
\ee

\subsection{Inverse See-Saw}
The inverse see-saw \cite{linear} is a variant of the double
see-saw. The texture of the neutral fermion mass matrix is the same as
in Eq.~\eqref{eq:exa1}, but now obeys the condition 
$M_S \ll m_D \ll m_{RS}$. In the limit of $M_S \rightarrow 0$ lepton
number is conserved and the scenario is natural in the 't Hooft sense 
\cite{Hooft}. It is the preferred scenario to arrange for
sizable unitarity violation. A recent discussion can be found in
Ref.~\cite{IMP}, and the results for $\eta$ in this paper agree.  

The calculation of the NLO terms proceeds in the same way as for the 
double see-saw, because we can perform the same replacement as in 
Eq.~\eqref{eqn:mathbbMDMR}: the eigenvalues of ${\mathbb M}_R$ (which
form a Pseudo-Dirac pair) are much larger than the entries in
$\mathbb M_D$. The effective light mass matrix $\tilde m_\nu$ and
the unitarity violating parameter $\eta$ look exactly as in 
Eq.~(\ref{eqn:lightdouble}) and Eq.~(\ref{eqn:etadouble}),
respectively. However, the term of order $M_S^2\,m_D^2/m_{RS}^4$ is
not anymore the dominating one, but can be neglected instead. 
This means that $\eta$ does basically not depend on $M_S$ and is given by 
\be 
\eta \simeq -\frac 12 m_D^\dagger \, (m_{RS}^\ast)^{-1} \, 
(m_{RS}^T)^{-1} \, m_D \, . 
\ee
With the suggestive values $m_D = 100$ GeV, $m_{RS} = 1$ TeV and $M_S = 0.1$
keV it follows that 
$\eta$ is of order $10^{-2}$. 
The leading term of the NLO correction to the mass matrix is  
\bea 
m_\nu^1 = -\frac 12\, 
 m_D^T \, m_{RS}^{-1} \left( M_S \, (m_{RS}^T)^{-1} \, m_D \, m_D^\dagger
\, (m_{RS}^\ast)^{-1} + \mbox{\small (last term)$^T$}\right) \, (m_{RS}^T)^{-1} \, m_D 
\, . 
\eea
It is of order $m_D^4 m_{RS}^{-4} \, M_S$ and for $m_D = 100$ GeV,
$m_{RS} = 1$ TeV and $M_S = 0.1$ keV  of order $10^{-2}$ eV. 
In analogy to the case treated in Section \ref{sec:app}, the sizable
NLO term has its origin in the moderate hierarchy of two sub-matrices
in the total neutral fermion mass matrix.

\subsection{Linear See-Saw}

The linear see-saw mechanism \cite{linear} arises when the neutral fermion 
mass matrix has the following form:
\be
{\cal M} = 
\left( 
\bad
0 & m_D^T & m_{DS}^T \\
m_D & 0 & m_{RS}^T \\
m_{DS} & m_{RS} & M_S \ea 
\right) . 
\ee
Here the non-zero $31$ entry $m_{DS}$ is assumed to be of weak scale, 
i.e.~of order $m_D$. In the following we will assume that $m_{RS}$ is
much larger than $m_D$ and $m_{DS}$. 
Its flavor structure may or may not be 
related to the flavor structure of $m_{DS}$. 
The relations between the other block matrices in 
$\cal M$ are that of the double or the inverse see-saw and are 
given above. We can introduce the notation 
\be
\mathbb M_D:=
\left(\ba
m_D\\
m_{DS}
\ea\right)\qquad \mathrm{and} \qquad \mathbb M_R:=\left(\baz
0&m^T_{RS}\\
m_{RS}&M_S\ea\right) ,
\ee
and write
\be\label{eqn:DoubleMass2}
\cal M = \left(
\baz 
0 & \mathbb M_D^T\\
\mathbb M_D & \mathbb M_R
\ea\right) .
\ee
The eigenvalues of $\mathbb M_R$ are much bigger than the entries 
in $\mathbb M_D$ which allows us again to apply 
the method of Section \ref{sec:GL} on matrix 
\eqref{eqn:DoubleMass2}. 
The uncorrected 
low energy mass matrix is easily obtained as 
\be\label{eqn:linear}
m_\nu^0 = m_D^T \, m_{RS}^{-1} \, M_S \, (m_{RS}^T)^{-1} \, 
m_D - \left[m_D^T \, m_{RS}^{-1} \, m_{DS}+ m_{DS}^T \, (m_{RS}^T)^{-1} \, 
m_D\right] . 
\ee
Note that if the first term was negligible and 
$m_{DS} \propto m_{RS}$, the flavor structure of 
$m_\nu$ is determined by the flavor structure of $m_D$. 
The unitarity violation $\eta$ is again determined by 
Eq.~\eqref{eq:etaB}. There are in total 5 different terms, two of
which are the known ones from Eq.~(\ref{eqn:etadouble}), and the
remaining three are 
\bea \label{eq:etalin}
\eta \simeq \eta^{\rm double} - \frac 12 \left( 
m_{DS}^\dagger \, (m_{RS}^\dagger)^{-1} \, m_{RS}^{-1} \, m_{DS} 
- m_D^\dagger \, (m_{RS}^\ast)^{-1} \, M_S^\ast \,
(m_{RS}^\dagger)^{-1} \, m_{RS}^{-1} \, m_{DS} \right. \\ 
\left. - m_{DS}^\dagger \, (m_{RS}^\dagger)^{-1} \, m_{RS}^{-1} \, M_S \, 
(m_{RS}^T)^{-1} \, m_D \right) \\
= {\cal O}
\left(
\frac{m_{DS}^2}{m_{RS}^2} \left( 
1 + \frac{m_D^2}{m_{DS}^2} + M_S \frac{m_D}{m_{DS} \, m_{RS}} 
+ M_S^2 \frac{m_D^2}{m_{RS}^2 \, m_{DS}^2} 
\right)
\right) . 
\eea
As mentioned above, quite often it holds in explicit realizations that 
$m_{DS} = \epsilon \, m_{RS}$, in which case 
the first new term in Eq.~(\ref{eq:etalin}) is proportional to
$\epsilon^2 \, \mathbbm{1}$, and the contribution to the mass matrix
is $- \epsilon \, (m_D + m_D^T)$. If we assume that the terms
containing $M_S$ are absent or sufficiently suppressed, then unitarity
violation is determined by terms of order $(m_{DS}^2 + m_D^2)
/m_{RS}^2$. Sizable violation of unitarity could be achieved if $m_D$ 
or $m_{DS}$ are sizable and not much smaller than $m_{RS}$.

\subsection{\label{sec:sing}Singular See-Saw}
Cases with a vanishing determinant of $M_R$ are called singular
see-saw \cite{sing0}, and have recently received some new attention
in the framework of sterile neutrino hints in LSND or MiniBooNE 
data \cite{sing_others}. We will shortly apply our approach to this case 
now.

In a three generation framework, the mass matrix is 
\be
{\cal M} = \left( 
\bas 
0 & 0 & 0 & a_1 & b_1 & c_1 \\
0 & 0 & 0 & a_2 & b_2 & c_2 \\
0 & 0 & 0 & a_3 & b_3 & c_3 \\
a_1 & a_2 & a_3 & 0 & 0 & 0 \\
b_1 & b_2 & b_3 & 0 & M_1 & 0 \\
c_1 & c_2 & c_3 & 0 & 0 & M_2 \\
\ea \right) . 
\ee 
There are two heavy mass states of order $M_{1,2}$, two light states of
order $m_D^2/M_R$ and two intermediate states of order $m_D$, which
form a Pseudo-Dirac pair. 
Realistic cases with 3 light neutrinos 
would require that $m_D$ and $M_R$ are $4 \times4$ matrices, the latter 
having rank 3. 

We can remove first the heavy states from
the discussion by identifying 
\be
\mathbb M_L = \left( 
\bav 
0 & 0 & 0 & a_1 \\
\cdot & 0 & 0 & a_2 \\
\cdot & \cdot & 0 & a_3 \\
\cdot & \cdot & \cdot & 0 
\ea \right)
~,~~
\mathbb M_D = 
\left( \bav 
b_1 & b_2 & b_3 & 0 \\
c_1 & c_2 & c_3 & 0 
\ea
\right)~,~~
\mathbb M_R = \left(\baz 
M_1 & 0 \\
0 & M_2 
\ea
\right).  
\ee
The low mass states (i.e., the small masses and the Pseudo-Dirac
pair) are obtained from diagonalizing 
\be
\mathbb M_L - \mathbb M_D^T \, \mathbb M_R^{-1} \, \mathbb M_D = 
 -\left( \bav 
\frac{b_1^2}{M_1} + \frac{c_1^2}{M_2} & 
\frac{b_1 \, b_2}{M_1} + \frac{c_1 \, c_2}{M_2} & 
\frac{b_1 \, b_3}{M_1} + \frac{c_1 \, c_3}{M_2} &
-a_1 \\ 
\cdot & \frac{b_2^2}{M_1} + \frac{c_2^2}{M_2} & 
\frac{b_2 \, b_3}{M_1} + \frac{c_2 \, c_3}{M_2} & 
-a_2 \\
\cdot & \cdot & \frac{b_3^2}{M_1} + \frac{c_3^2}{M_2} & 
-a_3 \\ 
\cdot & \cdot & \cdot & 0 
\ea \right) .
\ee
The corrections to this matrix can be evaluated using the expression
for the NLO term in case a triplet is present, see
Eq.~(\ref{eq:mlcor}). There are two terms, one steming from $\mathbb
M_R$, the other from $\mathbb M_L$. Their structure is different,
the contribution from $\mathbb M_R$ looks like 
\be
\left( 
\bav 
\ast & \ast & \ast & 0 \\
\ast & \ast & \ast & 0 \\
\ast & \ast & \ast & 0 \\
\ast & \ast & \ast & 0 \\
\ea
\right), 
\ee
where the non-zero entries are of order $m_D^4/M_R^3$. The
contribution from $\mathbb M_L$ has the structure 
\be
\left( 
\bav 
0 & 0 & 0 & \ast \\ 
0 & 0 & 0 & \ast \\ 
0 & 0 & 0 & \ast \\ 
0 & 0 & 0 & \ast \\ 
\ea
\right), 
\ee
where the non-zero entries are of order $m_D^3/M_R^2$. The relative
correction to all entries is therefore the same, namely of order 
$m_D^2/M_R^2$. 
As mentioned in Section \ref{sec:UnitarityViolation}, a triplet term 
does not contribute to unitarity
violation (as long as $m_L \ll M_R$), and $\eta$ is determined solely by 
$\mathbb M_D$ and $\mathbb M_R$. The result is 
\be
\eta = -\frac 12 
\left( 
\bav
\frac{|b_1|^2}{M_1^2} + \frac{|c_1|^2}{M_2^2} & 
\frac{b_1 \, b_2^\ast}{M_1^2} + \frac{c_1 \, c_2^\ast}{M_2^2} & 
\frac{b_1 \, b_3^\ast}{M_1^2} + \frac{c_1 \, c_3^\ast}{M_2^2} & 0 \\
\cdot & \frac{|b_2|^2}{M_1^2} + \frac{|c_2|^2}{M_2^2} & 
\frac{b_2 \, b_3^\ast}{M_1^2} + \frac{c_2 \, c_3^\ast}{M_2^2} & 0 \\
\cdot & \cdot & \frac{|b_3|^2}{M_1^2} + \frac{|c_3|^2}{M_2^2} & 0 \\
\cdot & \cdot & \cdot & 0\\
\ea
\right).
\ee
Though its entries are arbitrary, $\eta$ is effectively only a $3\times3$ 
matrix, having no effect for the fourth state, which is one of the 
Pseudo-Dirac states. \\

If for instance in the double see-saw of Section \ref{sec:double} the
matrix $m_{RS}$ was singular, we could now apply similar steps. After
suitable diagonalization of $\mathbb M_R$ in
Eq.~(\ref{eqn:mathbbMDMR}) it would (recall that $M_S \gg
m_{RS}$) take a form corresponding to 
diag$(0,m_{RS}^2/M_S,m_{RS}^2/M_S,M_S,M_S,M_S)$. Here the entries 
are understood as being of order $m_{RS}^2/M_S$ and $M_S$,
respectively. Because of $m_{RS}^2/M_S \gg m_D$ we can redefine
$\mathbb M_R$ as being a diagonal $5\times 5 $ matrix of the form 
diag$(m_{RS}^2/M_S,m_{RS}^2/M_S,M_S,M_S,M_S)$ and follow the procedure
of this subsection, finding a Pseudo-Dirac pair in the general case
etc. Interestingly, if $m_{RS}$ was rank 2, we could
write it as 
\be
m_{RS} = \left( 
\bad 
x_1 & x_2 & x_2/c \\
y_1 & y_2 & y_2/c \\
z_1 & z_2 & z_2/c 
\ea 
\right) . 
\ee
The eigenvector of the vanishing eigenvalue of $\mathbb M_R$ (which
has rank 5, if $M_S$ is non-singular) is proportional to 
$(0,-1/c,1,0,0,0)^T$, i.e.~a similar situation as for scaling treated
in Section \ref{sec:ss}. Note that with $m_{RS}$ having rank 2, and
$\mathbb M_R$ being rank 5, the full $9\times9$ mass matrix has rank
9; there is no vanishing eigenvalue.  
We will not discuss the cases of ``singular double
see-saw'' or ``singular inverse see-saw'' any further.

\section{\label{sec:concl}Conclusions and Summary}
With increasing precision in the experimental determination of
neutrino mass and lepton mixing parameters, care has to be taken in
giving theoretical predictions. In the present paper we have revisited
higher order corrections to the see-saw mechanism. The conventional
type I see-saw, as well as several popular variants were considered, and a
strategy to determine the next-to-leading order (NLO) terms was developed,
based on the well-known formalism for the type I see-saw. This can be
applied to determine both the NLO terms, as well as to obtain the
structure of the unitarity violation connected to see-saw mechanisms
with additional neutral fermions. Table \ref{tab:tab} summarizes the
structure of the zeroth and next-to-leading order terms, as well as
of the parameter describing the unitarity violation. 
We have identified situations in which no corrections arise to certain
observables, namely vanishing neutrino masses or $\mu$--$\tau$
symmetry.

While the standard type I see-saw 
implies insignificant NLO terms, there are cases with
phenomenologically interesting effects. 
This occurs when sub-matrices of the complete neutral fermion
mass matrix obey a moderate hierarchy (say, TeV and weak
scale). Examples are 
scenarios which explain the smallness of neutrino masses through
cancellations of large terms, or inverse see-saw frameworks. 
NLO corrections in the percent regime can arise in those cases. 

\newpage

\begin{center}
{\bf Acknowledgments}
\end{center}
This work is supported by the Deutsche Forschungsgemeinschaft (DFG) 
in the Transregio 27 ``Neutrinos and beyond -- weakly interacting 
particles in physics, astrophysics and cosmology''. W.R.~is supported
by the DFG in the project RO 2516/4-1 and by the ERC 
under the Starting Grant MANITOP.


\newpage

{\tiny\thispagestyle{empty}
\begin{sidewaystable}[h]\hspace{-2cm}\thispagestyle{empty}
\begin{tabular}{|c|c|c|c|}\hline 
& $m_\nu^0$ & $m_\nu^1$ & $\eta$ \\ \hline \hline 
type I & $ \left(\frac{m_D}{10^2 \, \rm GeV}
\right)^2 \left(\frac{10^{13} \, \rm GeV}{M_R} \right)$ eV 
& $10^{-22}$ $\left(\frac{m_D}{10^2 \, \rm GeV}
\right)^4 \left(\frac{10^{13} \, \rm GeV}{M_R} \right)^3$  eV  &
$10^{-22}$ $\left(\frac{m_D}{10^2 \, \rm GeV}
\right)^2 \left(\frac{10^{13} \, \rm GeV}{M_R} \right)^2 $ \\ \hline 
double & $\left(\frac{m_D}{10^2 \, \rm GeV}
\right)^2 \left(\frac{10^{16} \, \rm GeV}{m_{RS}} \right)^2 
\left(\frac{M_S}{10^{19} \, \rm GeV} \right)$ eV & 
$10^{-22}$ $ \left(\frac{m_D}{10^2 \, \rm GeV}
\right)^4 \left(\frac{10^{16} \, \rm GeV}{m_{RS}} \right)^6 
\left(\frac{M_S}{10^{19} \, \rm GeV} \right)^3$ eV 
& $10^{-22}$ $\left(\frac{m_D}{10^2 \, \rm GeV}
\right)^2 \left(\frac{10^{16} \, \rm GeV}{m_{RS}} \right)^4 
\left(\frac{M_S}{10^{19} \, \rm GeV} \right)^2$  
\\ \hline 
inverse & $\left(\frac{m_D}{10^2 \, \rm GeV}
\right)^2 \left(\frac{\rm TeV}{m_{RS}} \right)^2 
\left(\frac{M_S}{0.1 \, \rm keV} \right)$ eV & 
$10^{-2}$ $\left(\frac{m_D}{10^2 \, \rm GeV}
\right)^4 \left(\frac{\rm TeV}{m_{RS}} \right)^4 
\left(\frac{M_S}{0.1 \, \rm keV} \right)$  eV 
& $10^{-2}$ $\left(\frac{m_D}{10^2 \, \rm GeV}
\right)^2 \left(\frac{\rm TeV}{m_{RS}} \right)^2$ 
 \\ \hline 
linear & $\left(\frac{m_D}{10^2 \, \rm GeV} \right) 
\left(\frac{m_{DS}}{10^2 \, \rm GeV}  \right) 
\left(\frac{10^{13} \, \rm GeV}{m_{RS}} \right) $ eV & 
$10^{-22}$ $ \left( 
\left(\frac{m_D}{10^2 \, \rm GeV} \right)^{3~\rm or~1}  
\left(\frac{m_{DS}}{10^2 \, \rm GeV}  \right)^{1~\rm or~3}   
\right) \left(\frac{10^{13} \, \rm GeV}{m_{RS}} \right)^3
$ eV
& 
$10^{-22}$ $\left( \left(\frac{m_D}{10^2 \, \rm GeV}
\right)^2 + \left(\frac{m_{DS}}{10^2 \, \rm GeV} \right)^2 \right) 
\left(\frac{10^{13} \, \rm GeV}{m_{RS}} \right)^2$
\\ \hline
\end{tabular}
\caption{\label{tab:tab}
See-saw variants and their ``typical'' orders
of magnitude for the zeroth order mass matrix $m_\nu^0$, the NLO term 
$m_\nu^1$ and the unitarity violating  parameter $\eta$.}
\end{sidewaystable}\thispagestyle{empty}
}

\thispagestyle{empty}

\end{document}